\documentclass[twocolumn,aps,prl,showpacs]{revtex4}

\usepackage{graphicx}
\usepackage{amsmath}
\usepackage{amssymb}

\newcommand{\eref}[1]{(\ref{#1})}

\newcommand{\nnn}{\nonumber\\}
\newcommand{\PO}{\;.}   % after equation
\newcommand{\rme}{\mathrm{e}}
\newcommand{\rmi}{\mathrm{i}}
\newcommand{\rmd}{\mathrm{d}}

\newcommand{\sinc}{\mathop{\mathrm{sinc}}\nolimits}
\newcommand{\Z}{{{\mathbb{Z}}}}
\newcommand{\kB}{{k}_{\rm B}}

\newcommand{\LT}{L_\lambda}
\newcommand{\vm}{v_{\rm m}}
\newcommand{\vg}{v_{\rm g}}
\newcommand{\vgt}{\tilde{v}_{\rm g}}
\newcommand{\mg}{m_{\rm g}}
\newcommand{\pv}{p_{0}}
\newcommand{\Deltar}{\Delta r}
\newcommand{\wasell}{\ell}

\begin{document}

\title{Collisional Decoherence Observed in Matter Wave Interferometry}
\author{Klaus Hornberger}
\author{Stefan Uttenthaler}
\author{Bj\"{o}rn Brezger}
\author{Lucia Hackerm\"{u}ller}
\author{Markus Arndt}
\author{Anton Zeilinger}
%\email[]{zeilinger-office@exp.univie.ac.at}
\homepage[]{www.quantum.univie.ac.at}
\affiliation{Universit\"{a}t Wien, Institut f\"{u}r Experimentalphysik,
Boltzmanngasse 5, A-1090 Wien, Austria}

\date{March 14, 2003}

\begin{abstract}
We study the loss of spatial coherence in the
extended wave function of fullerenes due to collisions with background gases.
From the gradual suppression of quantum
interference with increasing gas pressure we are able to
support quantitatively both the predictions of decoherence theory
and our picture of the interaction process. We thus explore the
practical limits of matter wave interferometry at finite
gas pressures and estimate the required experimental vacuum
conditions for interferometry with even larger objects.
\end{abstract}

%03.65.-w   Quantum mechanics
%03.65.Ud   Entanglement and quantum nonlocality (e.g. EPR paradox,
%           Bell's inequalities, GHZ states, etc.)
%03.65.Ta   Foundations of quantum mechanics; measurement theory
%03.65.Yz   Decoherence; open systems; quantum statistical methods
%03.75.-b   Matter waves
%03.75.Dg   Atom and neutron interferometry
%39.20.+q   Atom interferometry techniques(see also 03.75.-b Matter waves,
%           and 03.75.Dg Atom and neutron interferometry in quantum mechanics)
\maketitle

Matter wave interferometers are based on \emph{quantum}
superpositions of spatially separated states of a single particle.
However, as is well known, the concept of wave-particle duality does
not apply to a \emph{classical} object which by definition never
occupies macroscopically distinct states simultaneously.
By performing interference experiments with particles of increasing complexity
one can therefore probe the borderline between these
incompatible  descriptions.

It is still a matter of debate how to explain the quantum-to-classical
transition in a unified framework.
Some theories contain an element beyond the
unitary evolution of quantum mechanics~\cite{Ghirardi1986a,Penrose1996a}
 --- which includes the `collapse' of the wave function as taught in many standard
textbooks.
Decoherence theory, on the other hand, remains
within the framework of the quantum theory \cite{Joos1985a,Zurek1991a,Tegmark1993a}.
It explains the decay of quantum coherences as being caused by the interaction of
the quantum object with its environment.

So far,
several decoherence experiments in atom interferometry
focused on the loss of coherence due to scattering of a
single~\cite{Pfau1994a,Chapman1995a} or a
few~\cite{Kokorowski2001a} laser photons by an atom. Other
authors proposed or realized schemes to encode which-path
information in  internal atomic degrees of freedom, thereby reducing the
interference contrast as well, in spite of a negligible change
in the atomic center-of-mass state~\cite{Scully1991a,Durr1998a}.
These studies are complemented by experiments which
quantitatively followed the decoherence of a coherent photon
state in a high-finesse microwave cavity~\cite{Brune1996a} or
of the motional state of a trapped ion~\cite{Myatt2000a}. However,  all
these experiments worked with few-level systems
and engineered environments.

In the present letter we quantitatively investigate a mechanism
which seems to be among the most natural and  most
effective sources of decoherence  in our macroscopic world, namely
collisions with gas particles.
From the controlled suppression of quantum interference
as a function of the gas pressure
we are able to
test both the predictions of
decoherence theory and our picture of the collisional interaction.

\begin{figure}[tb]
  \centering
  \includegraphics[bb= 6 278 581 591,width=0.9\columnwidth]{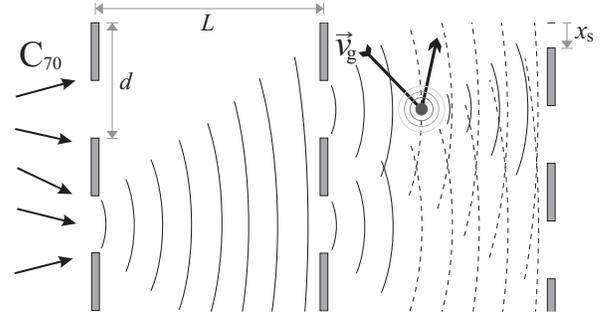}
  \caption{Schematic setup of the near-field interferometer for C$_{70}$ fullerenes.
    The third grating uncovers the interference pattern by yielding an
    oscillatory transmission with lateral shift $x_{\rm s}$.
    Collisions with gas molecules localize the
    molecular center-of-mass wave function leading to a reduced visibility
    of the interference pattern.
  }
  \label{fig:sketch}
\end{figure}

We note that the effect of atomic collisions in an atom interferometer was already
investigated  in~\cite{Schmiedmayer1995a}.  However,
decoherence effects were not observed in these experiments, since
the \emph{detected} atoms did not change the state of the
colliding gas sufficiently to leave behind the required path information for decoherence.
In contrast to that, our experiment uses \emph{massive}
C$_{70}$-fullerene molecules, and is based on a Talbot-Lau
interferometer (TLI) with a \emph{wide} acceptance angle.
Consequently, a fullerene molecule still enters the detector after
a typical collision, while the gas particle is
left in a state distinguishing the path taken.

Recently, the theoretically optimal interference contrast could be
observed in our high-vacuum TLI~\cite{Brezger2002a},
in spite of the high mass, temperature and complexity of the
fullerenes. This permits us to study now the
gradual loss of interference with increasing background gas pressure.
The central part of the experiment is sketched in Fig.~\ref{fig:sketch}. An
uncollimated, thermal beam of C$_{70}$ fullerenes passes three
identical vertical gold gratings, with a grating period of
d=991\,nm and a slit width of 475\,nm. They are separated by
an equal distance of $L=0.22$\,m which is the Talbot length
$\LT\equiv d^2/\lambda$ for molecules with a
velocity of 106  m/s (corresponding to a de Broglie wave length of $\lambda=4.46$\,pm).
A horizontal laser beam behind the
third grating ionizes the molecules regardless of their
horizontal position. Three height constrictions ---  the
oven orifice, the laser beam, and a horizontal slit half-way between
 --- determine the parabolic trajectories in the gravitational field and thus select
a narrow velocity distribution ($\Delta v/v = 8 \%$) out of
the molecular beam.

\begin{figure}[tb]
  \centering
  \includegraphics[width=\columnwidth]{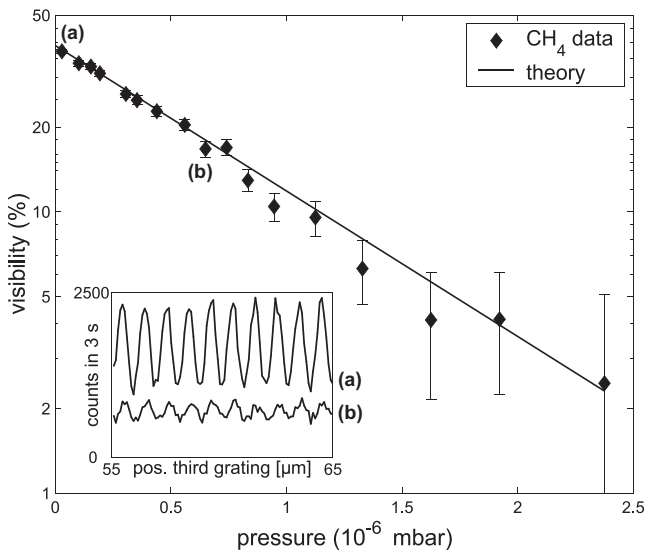}
  \caption{Fullerene fringe visibility
  vs.~methane   gas pressure on a semi-logarithmic scale.
  The exponential decay indicates that each collision leads to a
  complete loss of coherence. The solid line gives the prediction
  of decoherence theory, see
  text.  The inset shows the observed interference
  pattern at (a) $p=0.05\times 10^{-6}\,$mbar and (b) $p=0.6\times
  10^{-6}\,$mbar.}
  \label{fig:vp}
\end{figure}

The TLI
is based on
a near-field interference phenomenon,
the Talbot-Lau effect.  For a specific molecular wave length
it generates a high-contrast density pattern at the position of the
third grating, which is an image of the second one.
The quantum interferogram is then recorded by counting the number of
laser ionized fullerenes as a function of the lateral
position of the third grating $x_{\rm s}$.
Quantum mechanics predicts a transmission
periodic in $x_{\rm s}$ with period $d$
(see Fig.~\ref{fig:vp}),
\begin{equation}
\label{eq:transmission}
  T(x_{\rm s})  =
  \sum_{\wasell\in\Z}
  T_\wasell
  \exp\left(2\pi\rmi \wasell\frac{x_{\rm s}}{d}\right)
  \PO
\end{equation}
The Fourier coefficients $T_\wasell$ depend strongly on the molecular de
Broglie wavelength $\lambda$ (for $\wasell\neq 0$) as given in
\cite{BrezgerSubmitted}. They are determined by the grating configuration
and include the attractive Casimir-Polder interaction between the
fullerenes and the gold gratings.
From the observed $\lambda$-dependence of the
high-vacuum fringe visibility  we find that the signal is certainly caused
by near-field quantum interference, and not by  classical
dynamics~\cite{Brezger2002a,BrezgerSubmitted}.

In the following we use the TLI as a means of
monitoring the evolution of an extended, partially
coherent quantum state of the molecular center-of-mass.
The interaction with gas particles is  examined by filling the vacuum chamber
with various gases at low pressure (${p\leq 2.5\times 10^{-6}\,{\rm mbar}}$)
and room temperature.

In order to relate the expected loss of interference to
decoherence theory
\cite{Joos1985a,Zurek1991a,Gallis1990a,Altenmuller1997a} we
define the {\em decoherence function} $\eta$ as a factor to
the density matrix of the molecular center-of-mass state
$\rho_0(\mathbf{r},\mathbf{r}')$. It
describes the loss of coherence, i.e, the reduction of the  off-diagonal
elements in position representation, after one scattering event,
\begin{equation}
\label{eq:rho}
\rho(\mathbf{r},\mathbf{r}')=\rho_0(\mathbf{r},
\mathbf{r}')\,
\eta(|\mathbf{r}-\mathbf{r}'|)\PO
\end{equation}
This form follows from a
trace over the scattered gas particle.
It implies that the mass of the incident particle is much smaller
than the fullerene mass and that the distribution of the incoming
velocities is isotropic.

The  reduction of  interference is obtained by evolving the
molecular state from the first to the third grating subject to \eref{eq:rho}
which is equivalent to solving the corresponding
master equation in paraxial approximation.
While a detailed derivation will be given elsewhere, it is sufficient to note that the
final effect of collisional decoherence
is described completely
by a modification of the
Fourier coefficients,
\begin{equation}
\label{eq:Tdeco} T_\wasell\to T_\wasell
\exp\Big(-2n\sigma_{\rm eff} \int_0^L
\Big[1-\eta\Big(\wasell\frac{z\, \lambda}{d} \Big)\Big]\rmd
z\Big)
\PO
\end{equation}
Here, $n$ is the density of the gas environment and $\sigma_{\rm eff}$
the {effective total cross section} which accounts
also
for the thermal velocity distribution $g(\vg)$ in the gas.
We note that the component $T_0$ is left unchanged by
\eref{eq:Tdeco}, since $\eta(0)=1$ as required
from the conservation of probability in \eref{eq:rho}.
It follows that the average transmission remains constant, i.e,
the equation  describes the decoherence induced
by the gas, but no losses.

In order to obtain a kinematic interpretation of  \eref{eq:Tdeco}  we first
discuss the specific form of the decoherence function $\eta$ for large molecular masses.
By extending the
analysis in \cite{Joos1985a,Gallis1990a,Tegmark1993a}
and assuming an isotropic interaction potential, described by
the scattering amplitude  $f$,
we find
\begin{eqnarray}
\label{eq:eta}
\eta(\Deltar)&=& \int_0^{\infty}
\!\rmd \vg \;\frac{g(\vg )}{\sigma(\vg )}
\int\rmd\Omega
\left|f\big(\cos(\theta)\big)\right|^2
\nnn
&&
\times
\sinc\!\Big(\sin\!\Big(\frac{\theta}{2}\Big)\frac{2 \mg \vg \Deltar}{\hbar} \Big)
\PO
\end{eqnarray}
Here, the second integral covers
the scattering angles of the gas particle.
For  $\Deltar=|\mathbf{r}-\mathbf{r'}| \to 0$ it yields the total cross section
$\sigma(\vg)$ and we retrieve $\eta(0)=1$.
At finite separations $\Deltar$
the $\sinc$ function reduces
the contributions of the scattering amplitude
as the deflection angle $\theta$ grows,
i.e.\ with an increasing  momentum transfer during the collision.
The relevant length scale is set by the reciprocal momentum transfer in
units of $\hbar$.

Compare this to the  coherence in the molecular state which is
needed to contribute to the $\wasell$th Fourier component of the
signal. In order to illuminate coherently a region of size
$d/\wasell$ on the third grating from a distance $z$ the required
correlation in momentum must have a scale $\delta p$ with $\delta
p/p\times z\simeq d/\wasell$. Hence, $2\pi\hbar/\delta p\simeq
\wasell z\lambda /d$ which motivates the form of \eref{eq:Tdeco}:
Whenever the momentum kick experienced by the molecule at a
distance $z$ is large enough to destroy the required correlations
the molecule will not contribute to the interference signal. The
integration in \eref{eq:Tdeco} covers all scattering positions
between the second and the third grating, and by symmetry also
those between the first and the second one (yielding the factor
$2$).

The case of weak decoherence, where a single event yields only partial
which-way  information, was studied
in~\cite{Pfau1994a,Chapman1995a,Kokorowski2001a}.
There the
spatial coherence
in the center-of-mass state had a smaller scale
than the resolution set by the average momentum kick.
The present experiment explores the opposite regime since the
relevant path separations $\Delta r=\wasell z \lambda/d$ are by
orders of magnitude larger than the average reciprocal momentum
kick for almost all $z$. It follows that
the relevant long range coherences are destroyed completely
and independently of the separation $\Deltar$. This simplifies the
integration in \eref{eq:Tdeco} since we can now set
$\eta=0$ for $\ell\neq 0$
implying a localization rate which is determined only by the
effective cross section.
Since the fringe
contrast of our experiment is essentially determined by the basic
Fourier component $T_1$ \cite{BrezgerSubmitted} we thus expect a
visibility of
\begin{eqnarray}
\label{eq:Vdeco}
V(p)=
2 \frac{|T_1|}{T_0}
\exp\left(-\frac{2L\sigma_{\rm eff}}{\kB T}p\right)
=: V_0
 \,\rme^{-p/\pv}
\end{eqnarray}
as a function of the gas pressure $p=n\kB T$.
Note that although simple collisional loss
may lead to an exponential drop in count rate, loss alone will not affect the visibility.
The exponential decay of the fringe contrast
described by \eref{eq:Vdeco}
is a genuine effect of decoherence.

An experimental demonstration of collisional localization is
presented in Fig.~\ref{fig:vp}. It shows the pressure
dependence of the fringe visibility in the presence of methane
gas. A central molecular velocity of $\vm=117\,$m/s was chosen, corresponding to
a maximal vacuum visibility of $41\%$~\cite{Brezger2002a}.
The quantitative agreement with
our model (the solid line in Fig.~\ref{fig:vp}) is
obtained by extending the above reasoning by two additional points.

First, the momentum transfer is \emph{not} isotropic
in our experiment due to the directed motion of the molecules.
Nonetheless, any collision  localizes the molecule,
and the conclusion remains valid that the loss of coherence in
\eref{eq:Tdeco} is determined only by  $\sigma_{\rm eff}$. However,
in the effective cross section
the velocities of both the molecule $\vm$ and the gas must be taken into
account. Since the collisions
are governed by the isotropic London
dispersion force  \cite{Maitland1981a,Ruiz1997a}
they
are determined by
a single parameter $C_6$
(see Tab.~\ref{tab:C6}). Following  \cite{Maitland1981a} and after an integration over the
thermal distribution $g(\vg)$ we find
\begin{equation}
\label{eq:sigmaeff}
\sigma_{\rm eff}(\vm)=
\frac{C_6^{2/5}}{\hbar^{2/5}}
\,\frac{\vgt^{3/5}}{\vm}\,
\left(8.4946+1.6989\,\frac{\vm^2}{\vgt^2}\right)
%+\Or(\frac{\vm^4}{\vgt^4})
\end{equation}
with $\vgt$ the most probable velocity in the gas.
This expression, which is an asymptotic expansion for small
${\vm}/{\vgt}$,
predicts an effective cross section which exceeds the geometric one
by \emph{two} orders of magnitude.

\begin{table}
  \centering
\begin{tabular}{l@{\hspace{3ex}}c@{\hspace{9ex}}l@{\hspace{3ex}}c@{\hspace{9ex}}l@{\hspace{3ex}}c}
  % after \\ : \hline or \cline{col1-col2} \cline{col3-col4} ...
  gas &  $C_6$ &
  gas &  $C_6$ &
  gas &  $C_6$ \\
  \hline
  H$_2$ & 0.80
  &
  CH$_4$ & 3.3
  &
  Ar & 2.3
  \\
  D$_2$ & 0.77
  &
  N$_2$ & 2.1
  &
  Kr & 3.4
  \\
  He & 0.31
  &
  Ne & 0.71
  &
  Xe & 5.1
\end{tabular}

\caption{Van der Waals parameters for the interaction of C$_{70}$ fullerenes
  with various  gases, in units of
  ${\rm meV\,nm}^6$
  (obtained as outlined in \cite{Ruiz1997a} using data from
  \cite{C6sources}).%
  }
  \label{tab:C6}
\end{table}

The second point to be considered are the corrections
due to the particular constraints in our experiment, notably
the gravitational velocity selection and the finite size of the
detector. On the one hand, even those collisions which occur outside of the
interferometer can change the visibility since they  alter the
direction of the molecule
in the uncollimated beam,  which  reshuffles  the observed velocity
classes as a function of the gas pressure. On the other hand,
due to the finite width of the detector,
the observed molecules from the selected velocity class suffered
on average less collisions than the undetected ones.
In order to account for these effects  we  solve the
\emph{classical} phase space dynamics using a Monte Carlo method.
Our predictions for the visibility are then obtained by weighting
Eq. \eref{eq:Vdeco} with the classical velocity distribution
of those molecules which reach the detector.

As seen from Fig.~\ref{fig:vp} our calculation, which contains no
adjustable parameters, agrees well with the observed decrease of the visibility.
The simulation also  reproduces the
pressure dependence of the count rates and of the measured
velocity distributions.

\begin{figure}[b]
  \centering
  \includegraphics[width=0.9\columnwidth]{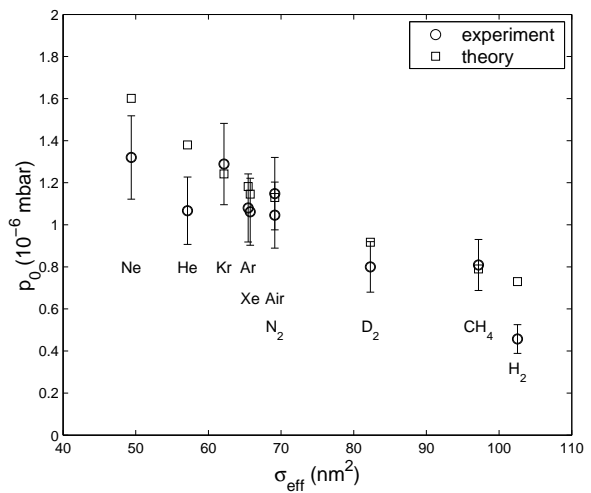}
  \caption{Experimental decoherence pressure $\pv$ for various gases compared to the predictions of
  decoherence theory.
  }
  \label{fig:druckfig}
\end{figure}

The loss of coherence with increasing pressure is
described by the ``decoherence pressure'' $\pv$
defined in \eref{eq:Vdeco}. It is determined experimentally by an
exponential fit to the pressure dependent visibilities as in Fig.~\ref{fig:vp}.
Figure~\ref{fig:druckfig} compares the measured
values of $\pv$ to the theoretical predictions for a number of mono-atomic
and molecular gases \footnote{The high-vacuum visibility
did not reach its optimal value in all of our measurements,
although it stayed significantly above the classical value.
We checked that the observed $p_0$  was independent of the attainable
maximum visibility, which is consistent with  attributing the visibility loss
to vibrational noise.}.
We find a very satisfactory agreement over the whole broad
range of masses and interaction strengths -- both of which cover almost two orders of
magnitude.
The experimental error is mainly due to the uncertainty in the
pressure measurement (about 15\%). The uncertainty of the
theoretical values amounts to about 5\% (not drawn in the figure) and is
related to lacking information about the velocity dependence of the
fullerene detection efficiency.

Most remarkable in Fig.~\ref{fig:druckfig} is
the weak dependence of the decoherence pressure on the specific type of collision partners.
This can be explained if we assume that the polarizability and therefore also $C_6$
are proportional to the mass of the scattered particle $\mg$. Then
\eref{eq:sigmaeff} shows that the mass
dependencies of the interaction constant and of the
mean gas velocity almost cancel out leaving $\sigma_{\rm
eff}\propto\mg^{1/10}$. This remaining dependence is so weak that the
deviations in the interaction constants  due to the particular electronic structure of the
gases outweigh the bulk behavior.
Xenon, for example, as the heaviest gas used,
lies right in the middle of the observed range of
decoherence pressures. We also note that the effect of molecular background gases
does not deviate systematically
from atomic ones.

Finally, based on the good overall agreement between experiment and theory we can
estimate the vacuum conditions that are required for the successful
observation of quantum interference of
much larger objects.
For the sake of an appealing example, let us consider a virus with a mass of $M=5\times 10^7\,$amu
interacting with molecular nitrogen (air) at room  temperature.
Since the static polarizability of large hydrocarbons
is closely proportional to their mass
the Slater-Kirkwood approximation~\cite[Chap.~13.3]{Hirschfelder1954a} for
the van der Waals parameter
yields $C_6(M$-N$_2)/{\rm meV nm}^6 \simeq 3.5\times 10^{-3}\,M$/amu.
Equations~\eref{eq:Vdeco} and \eref{eq:sigmaeff} then predict a decoherence pressure of
$\pv/{\rm mbar}\simeq 2.7\times 10^{-11}\,{\rm sec}\,\vm/L$.
By inserting $L=1\,$m  for the interferometer size and  $\vm=10\,$m/s for the velocity
we find that collisions would not limit quantum interference in a TL-interferometer even for an object
as large as a virus, provided we can reduce the background pressure to below
$p\simeq 3\times10^{-10}\,$mbar. This
is certainly feasible with available techniques.

In conclusion, our experiments investigate for the first time  the effect of decoherence due to
collisions with various gases. They are in very good quantitative agreement with decoherence theory.
While we are currently investigating other possible limits of matter wave interferometry
-- such as the emission of blackbody radiation -- it seems safe
to rephrase a famous word by R. Feynman~\cite{FeynmanTalk1959a}:
There is plenty of room at the \emph{top}.

\begin{acknowledgments}
This work has been supported by the European TMR network
(No.\ HPRN-CT-2000-00125),
the FWF projects F1505 and START Y177.
BB has been supported by
a EU Marie Curie fellowship (No.\ HPMF-CT-2000-00797),
and KH by the DFG Emmy Noether program.
\end{acknowledgments}

\end{document}